\documentclass[%
 reprint,
 amsmath,amssymb,
 aps,
citeautoscript,
prb,
]{revtex4-1}
\usepackage[colorlinks=false]{hyperref}
\usepackage{float}
\usepackage{graphicx}%
\usepackage{dcolumn}%
\usepackage{bm}%
\usepackage{xcolor}
\usepackage[normalem]{ulem}

\definecolor{commentColorTPM}{rgb}{0.8,0.0,0.2}
\definecolor{commentColorHS}{rgb}{0.0,0.2,0.6}
\definecolor{commentColorASN}{rgb}{0.2,0.6,0.0}

\newcommand{\m}[2]{{\mathbf{#1}_{#2}}}

\begin{document}

\preprint{APS/123-QED}

\title{Ultra-low-power second-order nonlinear optics on a chip}

\author{Timothy P. McKenna}
\thanks{These authors contributed equally to this work}
\author{Hubert S. Stokowski}%
\thanks{These authors contributed equally to this work}
\author{Vahid Ansari}%
\author{Jatadhari Mishra}%
\author{Marc Jankowski}%
\author{Christopher J. Sarabalis}%
\author{Jason F. Herrmann}%
\author{Carsten Langrock}%
\author{Martin M. Fejer}%
\author{Amir H. Safavi-Naeini}%
 \email{safavi@stanford.edu}
\affiliation{%
 Department of Applied Physics and Ginzton Laboratory, Stanford University\\
 348 Via Pueblo Mall, Stanford, California 94305, USA
}%

\date{\today}

\pacs{Valid PACS appear here}
\maketitle

{\bf
Second-order nonlinear optical processes are used to convert light from one wavelength to another and to generate quantum entanglement. Creating chip-scale devices to more efficiently realize and control these  interactions greatly increases the reach of photonics. Optical crystals and guided wave devices made from lithium niobate \cite{Weis1985} and potassium titanyl phosphate \cite{Bierlein1989} are typically used to realize second-order  processes \cite{Lim1989, Myers1995} but face significant drawbacks in scalability, power, and tailorability when compared to emerging integrated photonic systems. Silicon \cite{Jalali2006, Absil2015} or silicon nitride \cite{Romero-Garcia2013, Rahim2017} integrated photonic circuits enhance and control the third-order optical nonlinearity \cite{Moss2013,Gaeta2019} by confining light in dispersion-engineered waveguides and resonators. An analogous platform for second-order nonlinear optics remains an outstanding challenge in photonics. It would enable stronger interactions at lower power and reduce the number of competing nonlinear processes that emerge. Here we demonstrate efficient frequency doubling and parametric oscillation in a thin-film lithium niobate photonic circuit. Our device combines recent progress on periodically poled thin-film lithium niobate waveguides~\cite{Wang2018e,Jankowski2020,Rao2019,Chen2020} and low-loss microresonators~\cite{Lu2020b,Zhang2017a}.
Here we realize efficient ($>10\%$) second-harmonic generation and parametric oscillation with microwatts of optical power using a periodically-poled thin-film lithium niobate microresonator. The operating regimes of this system are controlled using the relative detuning of the intracavity resonances. During nondegenerate oscillation, the emission wavelength is tuned over terahertz by varying the pump frequency by 100’s of megahertz.
We observe highly-enhanced effective third-order nonlinearities caused by cascaded second-order processes resulting in parametric oscillation. These resonant second-order nonlinear circuits will form a crucial part of the emerging nonlinear and quantum photonics platforms.
}

The remarkable progress and impact of silicon photonics has led to the development of complex and high performance optical systems for communications, sensing, and quantum and classical information processing. In addition to linear passives, modulators, and detectors, many applications would significantly benefit from versatile nonlinearities. The lowest order nonlinearity of platforms like centrosymmetric silicon and amorphous silicon nitride is the third-order ($\chi^{(3)}$) nonlinearity, which has been used successfully for demonstrating optical frequency combs \cite{Shen2020,Okawachi2011a}, wavelength conversion \cite{Li2016}, and squeezed light generation \cite{Zhao2020b,Vaidya2020}. Efforts continue to further improve the efficiency and tailorability of these devices. One approach is to use a second-order ($\chi^{(2)}$) nonlinearity in an integrated device, by either breaking the symmetry of a crystal \cite{Timurdogan2017, Cazzanelli2016} or heterogeneously integrating a non-centrosymmetric material \cite{Chang2017, Weigel2016a}. 

\begin{figure*}[!] 
  \centering
  \includegraphics[width=\textwidth]{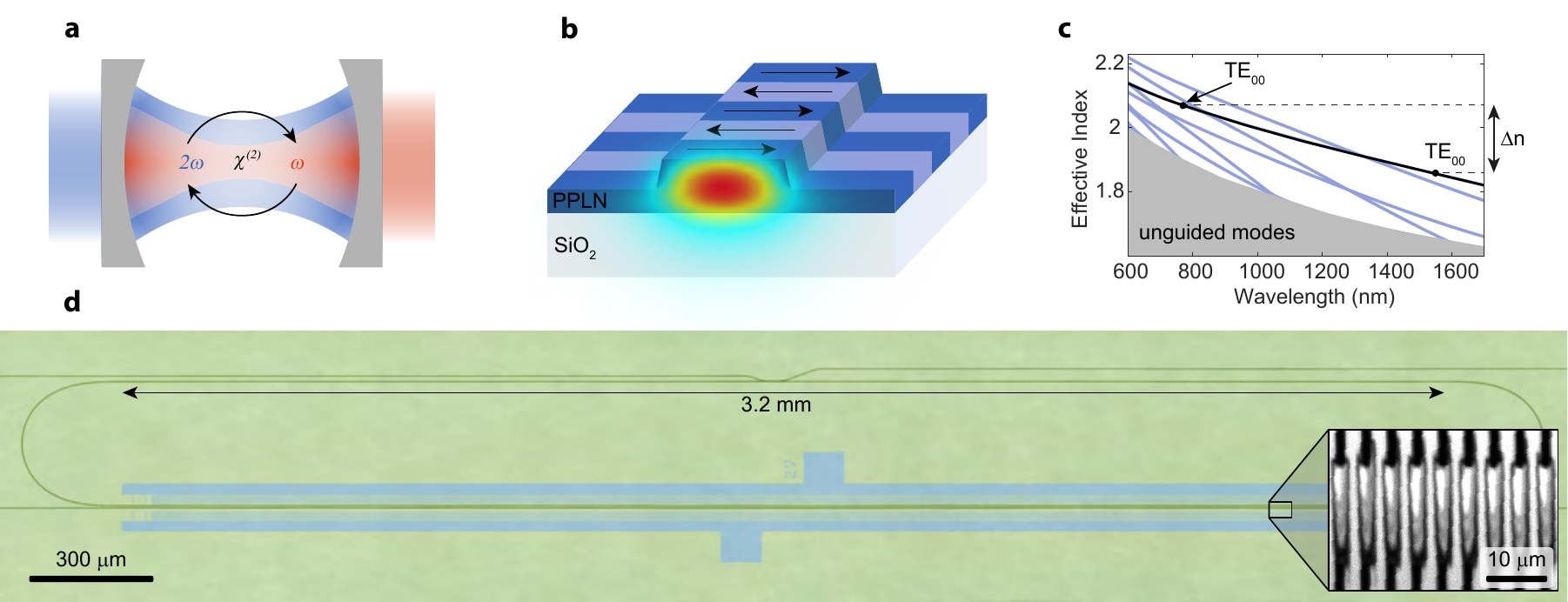}
\caption{\textbf{Integrated, resonant second-order nonlinear optical device.} \textbf{a}, Schematic of a resonant second-order nonlinear optical device. Driving the cavity with second harmonic light (blue) results in optical parametric oscillation at the fundamental; Driving at the fundamental frequency generates second harmonic light. \textbf{b}, Periodically-poled lithium niobate ridge waveguide that confines light to a small volume and supports nonlinear interactions. The transverse electric field parallel to the surface of the chip is plotted for the fundamental spatial mode. \textbf{c}, Effective index of the waveguide spatial modes as a function of wavelength. Periodic poling compensates for the phase velocity mismatch ($\propto \Delta n$) between fundamental and SH modes. \textbf{d}, The racetrack resonator used as a platform for nonlinear optics. Laser light is injected through an evanescent coupler on the top and undergoes nonlinear interaction in the bottom, periodically-poled section. The laser confocal microscope picture has been colorized; blue shading highlights the poling electrodes location during the fabrication process. Inset shows a second-harmonic microscope picture of the poled region. Inverted domains stretch between black electrode fingers.}
\label{fig:1}
\end{figure*}

Alternatively, photonic circuits may be built directly from a $\chi^{(2)}$ nonlinear material such as lithium niobate (LN).  In addition to supporting high-$Q$ optical resonances~\cite{Zhang2017a}, a large electro-optic coefficient~\cite{Wang2018d, Zhang2019, Li2020}, and Kerr nonlinearity~\cite{Wang2019, He2019}, LN can be periodically poled to compensate for phase mismatch due to dispersion~\cite{Jankowski2020, Wang2018e, Lu2020b, Rao2019, Chen2020}. Here we show ultra-efficient resonant $\chi^{(2)}$ nonlinear optical functions~(Fig.~\ref{fig:1}a) on a chip that incorporates quasi-phase-matching with a nonlinear optical resonator. 
\begin{figure}[!] 
  \centering
  \includegraphics[width=\columnwidth]{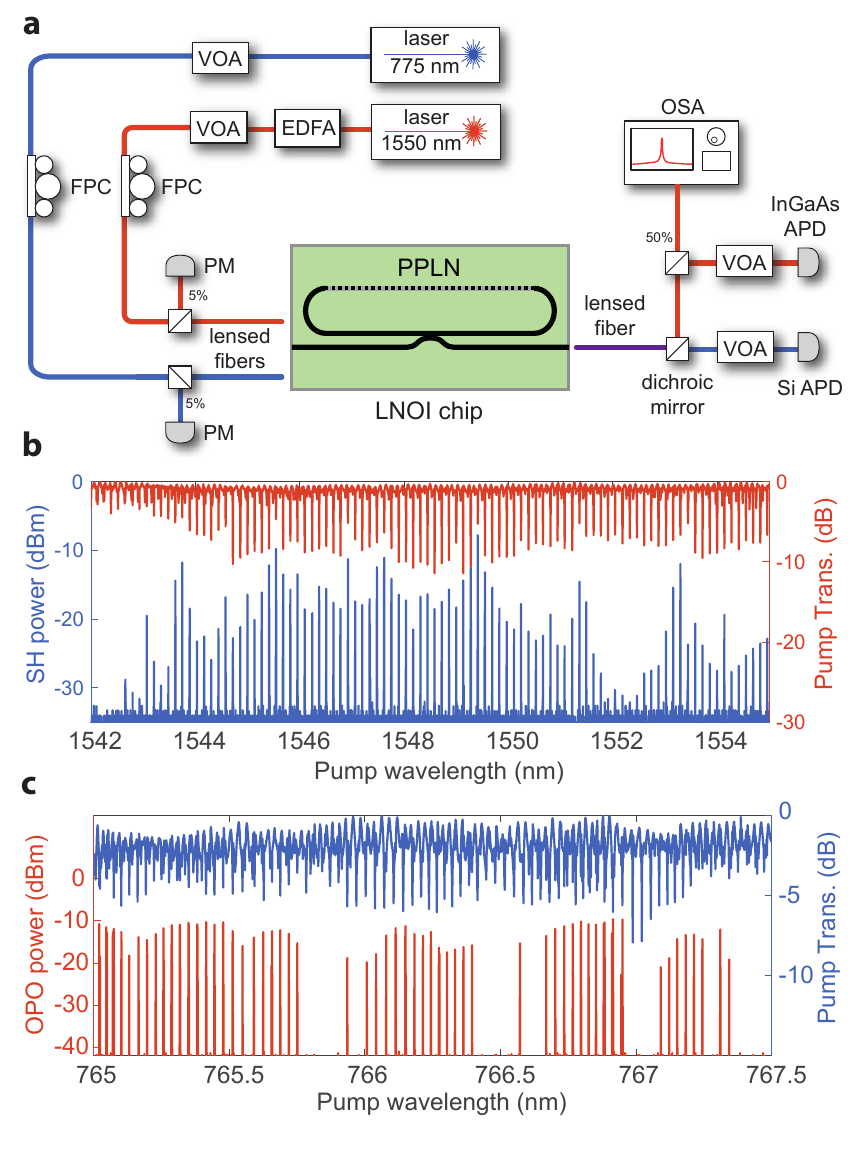}
\caption{\textbf{Characterization of Nonlinear Devices.}
\textbf{a}, Experimental setup. We couple light at around 775 or 1550~nm wavelengths onto the chip by aligning a lensed fiber to a cleaved edge facet to excite the coupling waveguide. Light is outcoupled from the chip and demultiplexed to detect the fundamental and second harmonic light separately. \textbf{b}, Scanning the near-infrared laser shows that second-harmonic generation occurs at wavelengths corresponding to modes of the resonator.  \textbf{c}, Scanning the blue pump laser across wavelength shows that many resonances surpass the parametric oscillation threshold. Abbreviations: VOA: variable optical attenuator, EDFA: erbium-doped fiber amplifier, FPC: fiber polarization controller, PM: power meter, OSA: optical spectrum analyzer, APD: avalanche photodiode.}
\label{fig:2}
\end{figure}

In this work, we make waveguides from a thin film of X-cut lithium niobate (Fig.~\ref{fig:1}b), which has its largest electro-optic and $\chi^{(2)}$ tensor components parallel to the surface of the chip. This orientation has been used in recent demonstrations of telecommunications modulators \cite{Wang2018d, Li2020}, frequency combs \cite{Zhang2019}, cryogenic frequency converters\cite{McKenna2020a, Holzgrafe2020a, Xu2020a}, and sources exhibiting quantum correlations \cite{Luo2017,Zhao2020a}, which form an emerging thin-film LN platform.  
We use magnesium oxide (MgO) doped lithium niobate to suppress pump-induced absorption and reduce the photorefractive damage typically experienced by devices fabricated with undoped congruently grown lithium niobate \cite{Furukawa1998}.  

Due to both its geometry and material properties, the dispersion of the waveguide introduces a phase velocity mismatch proportional to $\Delta n$ -- the difference in refractive indices between fundamental (FH) and second harmonic (SH) modes as shown in Figure \ref{fig:1}c. 
To achieve efficient nonlinear interactions, we compensate for the phase velocity mismatch by periodically poling the LN crystal. This quasi-phase-matching technique provides momentum conservation and enables the use of the same fundamental transverse electric (TE) spatial mode at both wavelengths~\cite{Wang2018e,Jankowski2020}. These modes exhibit the tightest confinement and have the strongest overlap with the large $d_{33}$ component of the $\chi^{(2)}$ nonlinear tensor, thereby enabling a large nonlinear interaction rate. We use a poling period of $\Lambda = \lambda_{\textrm{SH}}/\Delta n \approx 3.7$ $\mu$m. The inset of Figure \ref{fig:1}d shows a second-harmonic microscope picture of the periodic poling before waveguide fabrication. We observe the formation of oblong shapes with greyscale fringes between finger electrodes (black) that correspond to inverted crystal domains \cite{Rusing2019}.

The waveguide forms a racetrack resonator with a straight section length $L$ of 3.2~mm (see Fig. \ref{fig:1}d) that supports resonances across a broad range of wavelengths. Near the FH and SH frequencies, we measure intrinsic quality factors exceeding $10^6$, which dramatically enhance nonlinear processes by increasing the lifetimes of the interacting photons. 

The resonances at around the fundamental and second harmonic bands have frequencies $\omega_m$ and $\Omega_k$, with corresponding linewidths $\kappa_{A,m}$ and $\kappa_{B,k}$. We drive with pump frequency nearest to $\omega_0$ and $\Omega_0$ in following experiments. The FH mode frequencies vary with index as $\omega_m \approx \omega_0 + \zeta_1 m + \zeta_2  m^2/2$, where $\zeta_1$ is the free spectral range and $\zeta_2$ is a dispersion parameter. Temperature tuning of the devices changes the relative detuning between the modes and gives us fine control over the modal detuning $\mu\equiv\Omega_0 - 2\omega_0$. The small free spectral range of our device (17 GHz), allows us to tune $\mu$ while keeping the device within a few degrees of room temperature.

The $\chi^{(2)}$ optical nonlinearity of the material causes two FH resonances at $\omega_m$ and $\omega_n$, and the SH resonance at $\Omega_k$ to interact with each other at a rate $g_{k,nm}$. All of the dynamics of this system are captured by a set of coupled-mode equations for the fundamental ($A_m$) and second harmonic ($B_k$) field amplitudes. These amplitudes correspond to intracavity energies $\hbar\omega_m |A_m|^2$ and $\hbar\Omega_k |B_k|^2$, and evolve in time as
\begin{eqnarray}
\frac{\textrm{d}}{\textrm{d}t}  A_m &=& -\frac{\kappa_{A,m}}{2} A_m - 2i \sum_{k n} g_{k, n m} A_n^\ast B_k e^{-i\delta_{k,nm}t} \label{eqn:dAdt}\\
\frac{\textrm{d}}{\textrm{d}t} B_k &=& -\frac{\kappa_{B,k}}{2} B_k - i  \sum_{m n} g^{\ast}_{k, n m} A_m A_n e^{+i\delta_{k,nm}t} \label{eqn:dBdt},
\end{eqnarray}
with $\delta_{k,nm} \equiv \Omega_k - \omega_n - \omega_m$. To operate as an optical parametric oscillator (OPO), a laser driving term is added to the first equation, while adding a laser driving term to the second equation causes second harmonic generation (SHG) and eventually operation as a cascaded OPO.

Optical parametric oscillation occurs when the second-harmonic mode is driven to a sufficiently large steady-state cavity occupation $|B_0|^2$. The system will begin to oscillate at this input power, either as a degenerate OPO with emission into $\omega_0$ mode or as a nondegenerate OPO emitting into a pair of modes $\omega_{\pm m}$. The mode of oscillation is that with the lowest threshold $P_{\text{th},m}$, which strongly depends on laser detuning $\Delta$, modal detuning $\mu$, total loss $\kappa$, extrinsic loss $\kappa^{\text{(e)}}$, and dispersion $\zeta_2m^2$:
\begin{eqnarray}
P_{\text{th},m} &=& \frac{\hbar \Omega_b}{16 |g_{0,-mm}|^2}\frac{1 }{\kappa^{\text{(e)}}_{B,0}} \left( \Delta^2 + \left({\kappa_{B,0}}/{2} \right)^2 \right)\times\nonumber\\&&
\left(
\left( \Delta + \mu - \zeta_2 m^2 \right)^2 + \kappa_{A,m} \kappa_{A,-m}\right). \label{eqn:opo_P_th_l_mainText}
\end{eqnarray}
The pair of modes $\omega_{\pm m}$ with the lowest loss rates will experience the lowest threshold and oscillate first as we increase the pump power. Above the threshold, the OPO output power follows a square-root function of the input power $P_{B,0}$ provided that the input power is not sufficiently large to produce simultaneous oscillation of multiple mode pairs: 
\begin{eqnarray}
P_{\text{out}} = 
\cfrac{4\eta_{B,0}}{\Omega_{0}}
\left(
\eta_{A,m}\omega_{m}+ 
\eta_{A,-m}\omega_{-m}
\right)
\times \nonumber \\
P_{\text{th,m}}
\left(
\sqrt{\cfrac{P_{{B,0}}}{P_{\text{th,m}}}} - 1
\right).
\label{eqn:PoutOPOwithDisorder_main}
\end{eqnarray}
Here $\eta_{k,j}\equiv\kappa^{\text{(e)}}_{k,j}/\kappa_{k,j}$ is the cavity-waveguide coupling efficiency for $k \in \{A,B\}$ and $j$ being the index of a specific mode.

Driving the fundamental frequency $\omega_0$ generates light at the second harmonic mode $\Omega_0$. The efficiency of this process has a linear dependence on input power in the low power regime. Once the additional nonlinear conversion loss experienced by the FH mode (proportional to $8|g_{0,00} A_0|^2/\kappa_{B,0}$ with zero detuning) approaches the cavity linewidth $\kappa_{A,0}$, the cavity's effective coupling efficiency to the input light is reduced.  This leads to a sub-linear $P^{-1/3}$ dependence as the process now converts a substantial amount of pump photons to second harmonic in the resonator. A competing oscillation instability leading to parametric oscillations may prevent observing this power law. 

At high FH pump powers, the intracavity SH photon population at $\Omega_0$ is large enough to create an instability in the field amplitude of FH modes $A_m$, causing parametric oscillations when the generated SH intracavity photon number exceeds the threshold condition:
\begin{eqnarray}
| B_0|^2  \geq \frac{1}{16|g_{0,-mm}|^2  }&&\Big(\left(2\delta + \mu -\zeta_2 m^2\right)^2  \nonumber\\
&&+  \kappa_{A,m}\kappa_{A,-m} \Big). \label{eqn:copo_stab_crit_4}
\end{eqnarray}
We call this a cascaded OPO, since a cascade of two back-to-back $\chi^{(2)}$ processes leads to an effective $\chi^{(3)}$ process which is enhanced relative to that intrinsic to the material. The threshold for a cascaded OPO is a function of pump detuning $\delta$, modal detuning $\mu$, and dispersion $\zeta_2 m^2$. %

\begin{figure}[!] 
  \centering
  \includegraphics[width=\columnwidth]{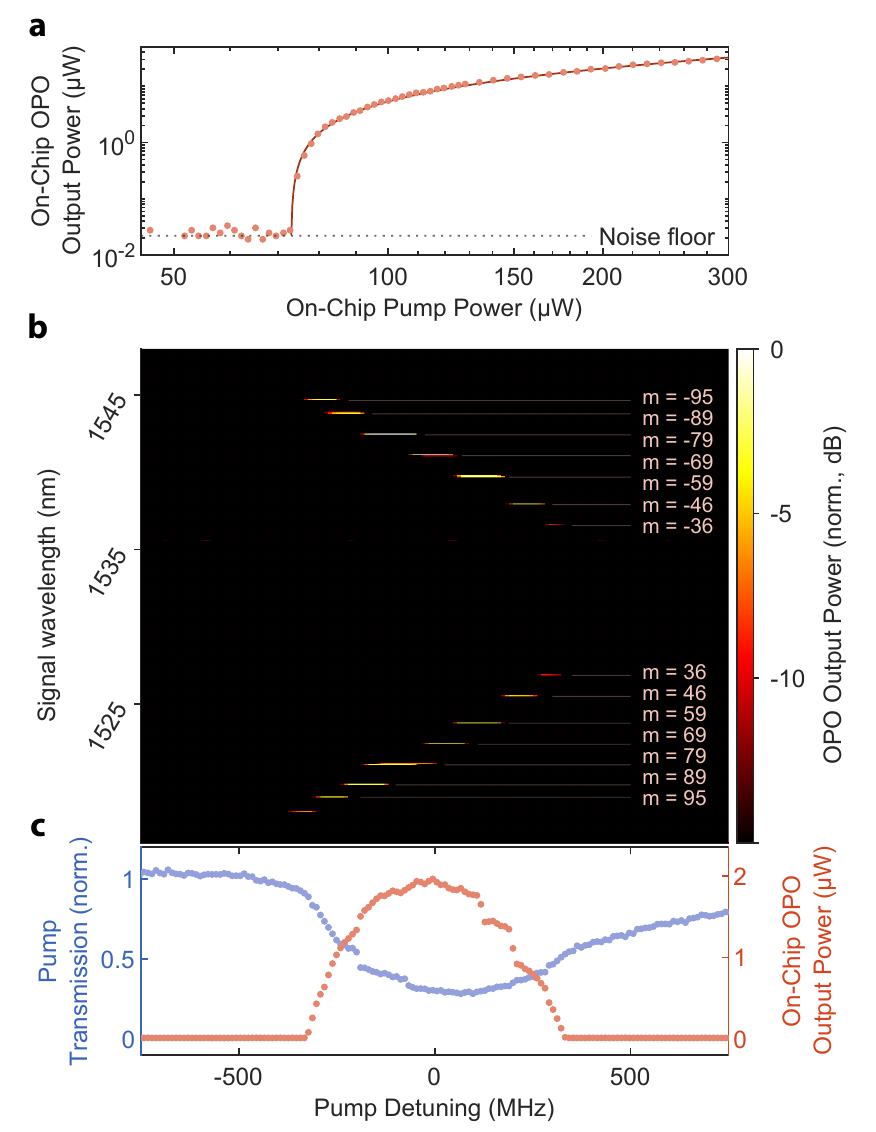}
\caption{\textbf{Optical Parametric Oscillation.} \textbf{a}, Threshold characterization of the OPO, we increase the 765.8~nm laser power until the oscillation begins at a threshold of around 73~$\mu\text{W}$; above it, the OPO power output follows a square-root relation. Dotted line represents detection noise floor. \textbf{b}, Tuning of non-degenerate OPO emission with pump wavelength, at a pump power of 250~$\mu\text{W}$. We can select OPO signal/idler pairs spanning 1537-1545~nm (1527-1519~nm) as the pump is swept over 650 MHz. \textbf{c}, SH resonance lineshape (blue points) aligned with the OPO response (red points) collected above threshold (250~$\mu\text{W}$) shows steps corresponding to switching between signal/idler pairs.}
\label{fig:3}
\end{figure}

\begin{figure*}[!] 
  \centering
  \includegraphics[width=\textwidth]{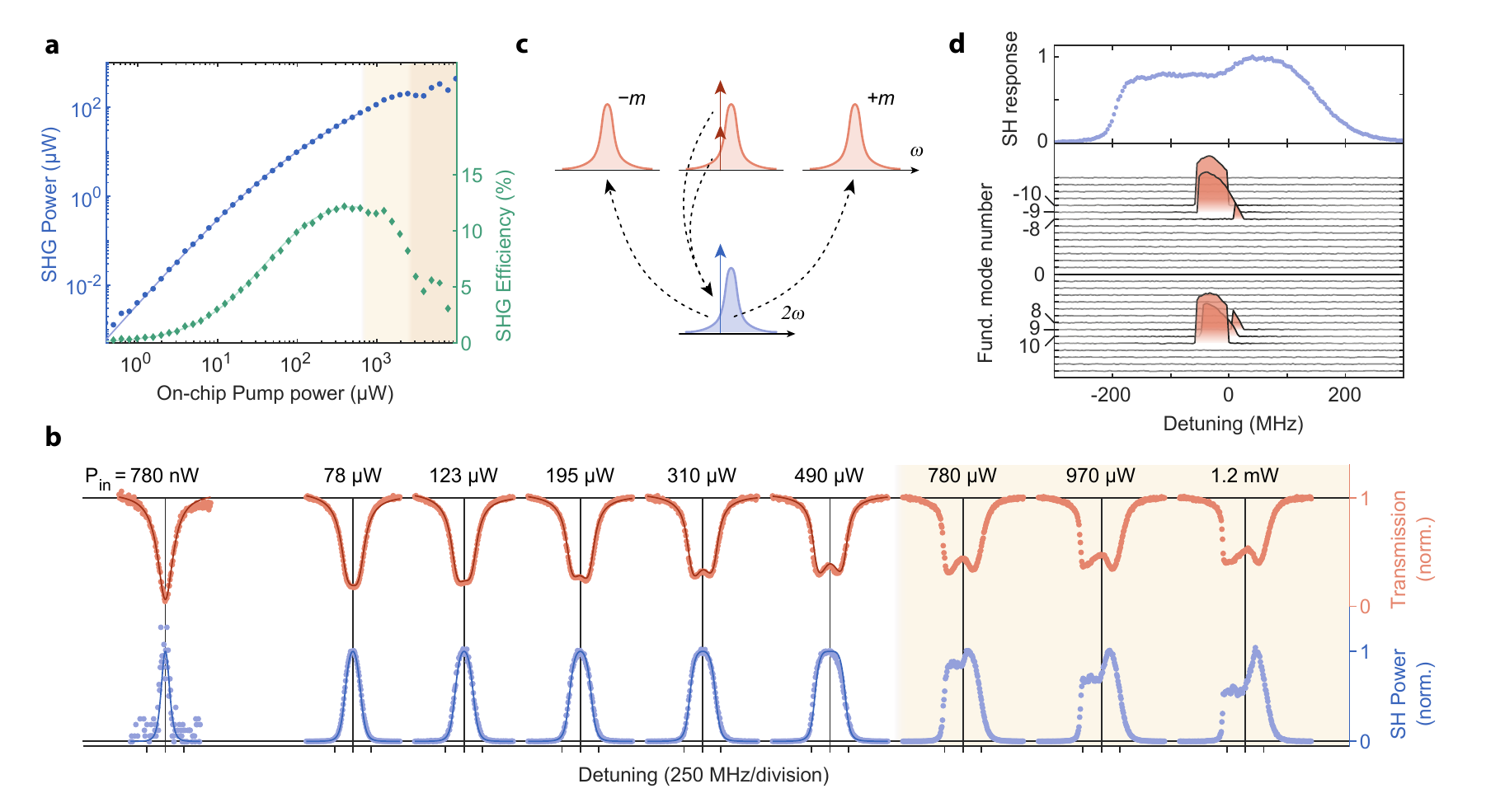}
\caption{\textbf{Second Harmonic Generation and Cascaded OPO.} \textbf{a}, 774.7~nm output power as a function of 1549.4~nm input power (blue circles, left axis) and SHG efficiency (green diamonds, right axis). Solid lines represent theoretical prediction. Light yellow shading corresponds to the region where we observe cascaded OPO. Darker shaded region have competing third-order nonlinear processes. \textbf{b}, Transmission (top) and SH (bottom) lineshapes evolve as a function of power. We plot theoretical curves (solid lines) on the top of data (red and blue points) up to the limit where the two-mode model breaks down and results in a cascaded OPO. A single parameter, the modal detuning $\mu$, is varied by about $0.04\kappa$ between fits. \textbf{c}, Cascaded OPO scheme - photons at the fundamental frequency drive the SHG process and create light at 2$\omega$. Sufficiently high power of the SH can drive the parametric oscillation back in the fundamental frequency range. \textbf{d}, Measured cascaded OPO, we observe light generation in modes symmetrically spaced from the pump frequency as the SH develops asymmetric lineshape shape (top panel).}
\label{fig:4}
\end{figure*}

We experimentally probe the nonlinear devices with the setup presented in Figure \ref{fig:2}a; we use two input paths to drive the resonator with fundamental and second harmonic frequency light -- shown in red and blue, respectively. We use the path connected to a tunable laser operating in telecommunication wavelengths to study the SHG and cascaded parametric oscillation processes. To drive a direct OPO, we use the input path connected to the shorter wavelength laser. The light is coupled into and out of the chip using lensed fibers. We separate the output light using a free-space setup with a dichroic mirror and send it to Si and InGaAs avalanche photodiodes. We show examples of transmission spectra and corresponding SHG and OPO signals in Figures~\ref{fig:2}b and \ref{fig:2}c, respectively. For spectrally-resolved measurements, we send part of the FH light to the optical spectrum analyzer. We calibrate the fiber-to-chip coupling efficiency based on power transmission measurements and fits of theoretical models to nonlinear response data (see Methods). The typical edge coupling efficiency across devices on the chip is 25-40\% at telecom wavelengths and 10-20\% for the second harmonic depending on fibers and alignment. For each experiment we measure these efficiencies to within less than a percent uncertainty (see Extended Data table I). All of the presented data refers to the on-chip power, accounting for the edge coupling loss.

We study the OPO by driving the device at around 765.8 nm and recording the generated light at close to twice the wavelength. We temperature tune the modal detuning $\mu$ close to zero to achieve degenerate operation (see Extended Data Figure 1). Given the modal detuning's temperature dependence and our device's comparatively small free-spectral range (about 17~GHz), we achieve an optimal operating point close to the room temperature, at $25.65^{\circ}$C. For the threshold measurement we detune $\mu$ from zero to allow the most efficient pair of modes at $\omega_{\pm m}$ to oscillate, following equation (\ref{eqn:opo_P_th_l_mainText}) (see also the condition defined by equation~(\ref{eqn:muCondition}) in the Methods section).
We plot the power of the generated near infrared light in Figure \ref{fig:3}a. The output power vs. input power curve reveals the threshold of oscillation around 73 $\mu\text{W}$, which we extract from fitting equation (\ref{eqn:PoutOPOwithDisorder_main}). A maximum efficiency of $11\%$ is measured.
Tuning the pump laser wavelength allows for effective selection for the frequencies of oscillating signal-idler pairs of modes. By changing the laser detuning $\Delta$, we observe seven different OPO wavelength pairs generated in the resonator. Figure \ref{fig:3}b shows the OPO emission spectrum as a function of pump detuning with a pump power of 250~$\mu\text{W}$. By tuning the pump laser by just 650 MHz, we can address signal modes across a band of over 1 THz. Figure \ref{fig:3}c shows the pump transmission and OPO emitted power a function of the pump detuning. Detunings of the pump laser relative to the SH cavity mode result in exciting different OPO modes. We can resolve steps on the transmission and OPO emitted power that correspond to switching between different operation modes.

To demonstrate second harmonic generation, we drive the FH mode at 1549.4 nm and measure the resulting frequency doubled light at the output. The device temperature is $30.5^\circ$C.
Fig.~\ref{fig:4}a shows the peak SH power generated as a function of input FH power. A maximum efficiency of 12\% is achieved with 390~$\mu\text{W}$ of input power in the feed waveguide, which agrees with the coupled mode theory (solid lines) that includes only the $A_0$ and $B_0$ fields. Fig. \ref{fig:4}b shows how the transmission lineshape and the SH response change as a function of pump power. 
As the pump power increases, the transmission lineshape widens and becomes shallower due to the additional two photon loss induced by the nonlinearity. At pump powers around 200~$\mu\text{W}$, the transmission lineshape forms two distinct valleys, consistent with our coupled mode theory simulations.

At higher input powers (the yellow shaded region of Fig. \ref{fig:4}b), the SH response becomes asymmetrical with a distinct drop in SH power for negative pump detunings, $\delta < 0$. At these powers, the intracavity SH light is intense enough to create an instability in the field amplitude of the fundamental modes at $\omega_{\pm m}$, causing parametric oscillations as visualized in Fig. \ref{fig:4}c. The cascade of two $\chi^{(2)}$ processes creates the parametric oscillation. The normal dispersion of the waveguide ($\zeta_2 < 0$) creates a lower threshold condition for negative pump detunings ($\delta<0$), see equation (\ref{eqn:copo_stab_crit_4}). The drop in SH output power at those laser detunings is because SH light at $\Omega_0$ converts back to FH power at $\omega_{\pm m}$.

We spectrally resolve the cascaded parametric oscillations as a function of laser detuning and confirm that the first sideband fundamental modes oscillate at a threshold of 690~$\mu\text{W}$ of on-chip pump power. Fig. \ref{fig:4}d shows multiple sideband oscillations that occur at a pump power of 930~$\mu\text{W}$. Particular signal-idler pairs oscillate as a function of pump laser detuning as expected from equation (\ref{eqn:copo_stab_crit_4}). Disorder in the mode spacing and quality factors causes certain mode pairs to oscillate before others, consistent with coupled mode simulations.

We expect to find ultra-efficient second-order nonlinear photonic circuits, such as the frequency doubler and parametric oscillator demonstrated in this work, in a number of emerging low-power and quantum applications in the near future. Together with the high performance integrated devices and components that are being developed for the thin-film LN platform, the promise of a new class of versatile integrated photonic technologies may soon be realized. In addition to sources of broadband and quantum light for sensing and communications, integrated ultra-low-power OPOs can be used for computation with coherent ising machines~\cite{Yamamoto2017} and cluster states \cite{Yokoyama2013a, Chen2014a}. 
\emph{Note:} In the final stages of preparing this manuscript we became aware of a demonstration of low power optical parametric oscillator in a lithium niobate microresonator~\cite{Lu2021}.

\bibstyle{naturemag}

\clearpage
\section*{Methods}

\subsection{Fabrication}
We fabricate all devices with X-cut thin-film lithium niobate-on-insulator (LNOI) wafers. The material consists of a 500 nm film of LN bonded to a 2 $\mu$m layer of silicon dioxide on top of an LN handle wafer. 

We pattern the optical devices using electron beam lithography (JEOL 6300-FS, 100-kV) and transfer the design to the LN via Argon ion milling. The waveguide width is 1.2 $\mu$m, and the etch depth is 300 nm which leaves a 200 nm slab of LN beneath the waveguide. We deposit 700 nm of PECVD silicon dioxide at a temperature of 350 $ ^{\circ}$C as cladding. 

 We perform the periodic poling step before waveguide fabrication. For periodic poling, we use electron-beam evaporated Cr electrodes with an electron beam lithography-based liftoff process and apply high-voltage pulses similar to Nagy \textit{et al.} \cite{Nagy2019} to invert the crystal domains. Upon completion of the poling, we remove the electrodes. 

Chip edge facet preparation is done using a DISCO DFL7340 laser saw. High energy pulses are focused into the substrate to create a periodic array of damage locations, which act as nucleation sites for crack propagation and result in a uniform and smooth cleave. 

\subsection{Experimental Setup}
We characterize fabricated devices in a simplified experimental setup shown in Figure \ref{fig:2}a. In the FH input path, we use SMF-28 fibers. 5\% of the laser light (Santec TSL-550, 1480-1630 nm) goes into a Mach-Zehnder interferometer (MZI) with an FSR of 67.7 MHz used to calibrate the relative wavelength during laser wavelength sweeps (not shown in \ref{fig:2}a). 95\% of the light goes to erbium-doped fiber amplifier (EDFA) with a fixed output power of 250~mW followed by a variable optical attenuator. Next, the light passes through a fiber polarization controller (FPC), and we tap 5\% of it just before the input lensed fiber for power calibration with a power meter (Newport 918D-IR-OD3R). The light then couples to the chip facet through an SMF-28 lensed fiber.

In the SH path, we use a Velocity TLB-6700 laser that operates in the 765-781 nm range. This entire path uses 780HP fiber to maintain single-mode operation. A 5\% tap outcouples part of the light an MZI with an FSR of 39.9 MHz to calibrate laser wavelength sweeps (not shown in \ref{fig:2}a).  A variable optical attenuator controls the remaining laser power, and we control the polarization with an FPC. 5\% of the light goes to a power meter (Newport 918D-SL-OD3R) for input power calibration, and we focus the rest of it on the chip facet through a 780HP lensed fiber.

Once the light exits the output edge facet of the chip, we collect it into a lensed SMF-28 fiber, similar to the one used in the FH input path. We outcouple the light into free space and demultiplex with a 1000 nm short pass dichroic mirror. After the dichroic mirror, SH and FH paths are additionally filtered to ensure no cross talk, and we detect SH and FH light with avalanche photodiodes (Thorlabs APD410A and Thorlabs APD410, respectively). Variable optical attenuators are used before the APDs to avoid saturation. We split 50\% of the FH light into an optical spectrum analyzer (OSA, Yokogawa AQ6370D) for spectrally-resolved measurements of the OPO.

We use different, but similar, devices on the same chip for the SHG and OPO experiments. The chip sits directly on a thermo-electric cooler for temperature adjustment.

\subsection{OPO Characterization}
We characterize all of the optical resonances that take part in the optical parametric oscillation using linear spectroscopy at powers substantially below nonlinear effects. For these measurements, we sweep the wavelength of tunable lasers in the FH and SH bands and fit the transmission dips with lorentzian lineshapes. We determine the total and intrinsic quality factors of the second harmonic mode to be $Q_{B,0} = 0.88 \times 10^{6}$ and $Q_{B,0}^{\text{(i)}} = 1.5 \times 10^{6}$, respectively. We find the quality factors of the OPO signal modes corresponding to the curve in \ref{fig:2}a to be: $Q_{A,m} = 0.68 \times 10^{6}$, $Q_{A,m}^{\text{(i)}} = 0.8 \times 10^{6}$, $Q_{A,-m} = 0.94 \times 10^{6}$, $Q_{A,-m}^{\text{(i)}} = 1.5 \times 10^{6}$. We perform an independent second harmonic generation measurements to determine if the FH and SH modes are under or overcoupled. The analysis of transmission lineshapes as a function of pump power confirms that all modes are undercoupled. From the determined threshold of 73 $\mu$W we deduce a coupling rate $|g_{0,-mm}|$ of 150 kHz which is close to the simulated value of 186 kHz.

We measure the input fiber-to-chip coupling with an independent transmission measurement using 780HP lensed fibers at the input and the output chip edges. We assume the input and output coupling is identical, an assumption based on experience with multiple devices on the chip used for the experiment, and find the input edge coupling efficiency to be 13\%. We extract the output fiber-to-chip coupling efficiency at the OPO wavelength by fitting the data in Fig. \ref{fig:3}a to $\eta_{\text{FH}}P_{\text{out}}$ using equation \ref{eqn:PoutOPOwithDisorder_main}. We infer $\eta_{FH}$ = 37\% coupling efficiency, which we confirm with an independent transmission measurement.

\subsection{SHG Characterization}
We characterize the modes contributing to the second harmonic generation in an analogous way to the OPO. From the Lorentzian fits at low power we find quality factors of $Q_{B,0} = 0.82 \times 10^{6}$, $Q_{B,0}^{\text{(i)}} = 1.2 \times 10^{6}$, $Q_{A,0} = 0.75 \times 10^{6}$, and $Q_{A,0}^{\text{(i)}} = 1.2 \times 10^{6}$. Moreover, we use a method for fitting nonlinear lineshapes at high power, as mentioned in the main text. For this purpose, we solve equations \ref{eqn:steadystate_A0} and \ref{eqn:steadystate_BK} numerically and fit the resulting curves as a function of detuning to the data. We use the ten lineshapes at the pump power between 80 and 620 $\mu$W, which allows us to observe changes due to the second-order nonlinearities but avoid the effects of the cascaded OPO. From this procedure we find average $Q_{A,0} = 0.74 \times 10^{6}$, and $Q_{A,0}^{\text{(i)}} = 1.2 \times 10^{6}$ and standard deviation of less than 4\% which agrees with the low power fit. From fitting nonlinear lineshapes, we also extract the coupling rate $|g_{0,00}|$ to be about 130 kHz, which agrees with our theoretical prediction of 170 kHz. In the main text, we use averaged values to plot the theoretical lineshapes and only vary the modal detuning to account for small temperature fluctuations. For the SHG device, we make transmission measurements and find the coupling efficiencies to be to be 26\% and 11\% at the FH and SH, respectively. 

We calculate the theoretical relationship between the pump power, SHG power, and SHG efficiency by numerically solving equations \ref{eqn:steadystate_A0} and \ref{eqn:steadystate_BK} for zero detuning. For the solid lines plotted in Figure \ref{fig:4}a, we use quality factors and the nonlinear coupling rate from the measurements described in the previous paragraph.

\subsection{Resolving OPO Lines}
We use an OSA (Yokogawa AQ370D) to characterize the frequency content of the OPO output spectrum as a function of pump laser detuning. With a constant pump power, we repeatedly sweep the laser wavelength across the SH resonance and record the SH and FH response with APDs (see section Experimental Setup). A portion of the generated FH light is detected by the OSA operating in zero-span mode with a 0.1 nm filter bandwidth, which is less than the approximately 0.135 nm free spectral range of the FH modes. We step the center wavelength of the OSA across a 40 nm span with a 50\% overlap in OSA filter spans. We record the detected power on the OSA synchronously with the APD detector voltages for each wavelength step. Repeated laser sweeps with different OSA filter center wavelengths produce a map of the OPO frequency content as a function of laser detuning shown in \ref{fig:3}b. 

To characterize the cascaded parametric oscillations as shown in \ref{fig:4}d, we first find every potential OPO line's precise location ($\omega_m$) by performing a broad sweep of the FH pump laser and record the resonance frequencies. We then proceeded with the measurement in an identical fashion to the standard OPO, but with the 0.1 nm wide OSA filters placed precisely at the FH mode locations without any overlap between filters.

\subsection{Coupled Mode Theory Equations}

The Hamiltonian of the system is used to find the equations of motion in the rotating frame. 
\begin{eqnarray}
\frac{\textrm{d}}{\textrm{d}t} A_m &=& -\frac{\kappa_{A,m}}{2} A_m - 2i \sum_{k n} g_{k,nm} A_n^\ast B_k e^{-i\delta_{k,nm}t} - \nonumber\\&&~~~~ \sqrt{\kappa^{\text{(e)}}_{A,m}}F_m e^{-i(\omega_L - \omega_m)t} \label{eqn:dAdt}\\
\frac{\textrm{d}}{\textrm{d}t}  B_k &=& -\frac{\kappa_{B,k}}{2} B_k - i  \sum_{m n} g^\ast_{k,nm} A_m A_n e^{+i\delta_{k,nm}t} -\nonumber\\&&~~~~ \sqrt{\kappa^{\text{(e)}}_{B,k}}G_k e^{-i(\Omega_L - \Omega_k)t} \label{eqn:dBdt},
\end{eqnarray}
with $\delta_{k,nm}=\Omega_k -\omega_n -\omega_m$. $A_m$ is the fundamental field amplitude at $\omega_m$, and $B_k$ is the second harmonic field amplitude at $\Omega_k$.

\subsection{Parametric Oscillation Theory}

We consider the case where the SH modes are driven at frequency $\Omega_L$ and the $A$ modes are not excited and calculate the stability criterion for the $A$ modes based on equation (\ref{eqn:dBdt}):
\begin{eqnarray}
\frac{\textrm{d}}{\textrm{d}t}  B_0 &=& -\frac{\kappa_{B,0}}{2} B_0 -  \sqrt{\kappa^{\text{(e)}}_{B,0}}G_0 e^{-i(\Omega_L - \Omega_0)t} \label{eqn:opo_dBdt}.
\end{eqnarray}
We go into a rotating frame $B_0 = \Tilde{B}_0 e^{-i\Delta t}$ with frequency $\Delta = \Omega_L - \Omega_0$ defined as the detuning between the laser drive and the $B$  mode, which we can solve in steady-state to obtain:
\begin{eqnarray}
\Tilde{B}_0 = \frac{ \sqrt{\kappa^{\text{(e)}}_{B,0}}G_0}{i\Delta - {\kappa_{B,0}}/{2}}.\label{eqn:opo_B0_driven}
\end{eqnarray}
We now consider two $A$ modes at frequencies $\omega_m$ and $\omega_{-m}$ which are coupled by the intracavity population of $B_0$. Their coupling leads to a pair of equations
\begin{eqnarray}
\frac{\textrm{d}}{\textrm{d}t}  A_m &=& -\frac{\kappa_{A,m}}{2} A_m - 2 i g_{0,-mm} A_{-m}^\ast B_0 e^{-i\delta_{0,-mm}t} \nonumber %
\\
\frac{\textrm{d}}{\textrm{d}t}  A^\ast_{-m} &=& -\frac{\kappa_{A,-m}}{2} A_{-m}^\ast + 2i g^\ast_{0,-mm} A_{m} B^\ast_0 e^{i\delta_{0,-mm}t} \nonumber%
\end{eqnarray}
which become unstable for sufficiently large $|B_0|$. To see this note that $\delta_{0,-mm} = (\Omega_0 - 2\omega_0) - \zeta_2 m^2$ allowing us to move into a rotating frame with 
\begin{eqnarray}
A_m = \Tilde{A}_m \exp\left( -i\frac{\Delta+\mu-\zeta_2 m^2}{2}t \right)
\end{eqnarray}
where $\mu\equiv\Omega_0 - 2\omega_0$ is the \emph{modal detuning} between the driven SH and closest FH mode, which in our experiment is set by tuning the temperature.
In this frame, the equations become time-independent, and we obtain the stability criterion (assuming $\kappa_{\pm m}$ are equal for simplicity):
\begin{eqnarray}
16|g_{0,-mm}|^2 |B_0|^2 \ge \left( \Delta + \mu - \zeta_2 m^2 \right)^2 + \left(\kappa_{A,m}\right)^2.\label{eqn:opo_stab_crit_1}
\end{eqnarray}
To relate this to the input photon flux at the SH frequency $\Omega_0$, we replace $B_0$ using eqn.~(\ref{eqn:opo_B0_driven}), to obtain
\begin{eqnarray}
16|g_{0,-mm}|^2&& |G_0|^2 \kappa^{\text{(e)}}_{B,0} \ge 
\left( \Delta^2 + \left({\kappa_{B,0}}/{2} \right)^2 \right)\times\nonumber\\&&~~~
\left(
\left( \Delta + \mu - \zeta_2 m^2 \right)^2 + \left({\kappa_{A,m}} \right)^2\right).\label{eqn:opo_stab_crit_2}
\end{eqnarray}
We can see from here that the lowest degenerate oscillation threshold can be achieved when $\mu=0$ and $\Delta=0$:
\begin{eqnarray}
4|g_{0,-mm}|^2 |G_0|^2 \kappa^{\text{(e)}}_{B,0} &\ge& 
\left({\kappa_{B,0}}/{2} \right)^2
 \left({\kappa_{A,m}}/{2} \right)^2~\text{or,}\\
 P_\text{th,0} &=&\frac{\hbar \Omega_0}{64 |g_{0,00}|^2}\frac{\kappa^{2}_{B,0} \kappa^{2}_{A,0} }{\kappa^{\text{(e)}}_{B,0}}.
\end{eqnarray}
More generally, the OPO will oscillate first in the mode $m$ for which $ P_{\text{th},m}$ is the lowest, where
\begin{eqnarray}
P_{\text{th},m} &=& \frac{\hbar \Omega_0}{16 |g_{0,-mm}|^2}\frac{1 }{\kappa^{\text{(e)}}_{B,0}} \left( \Delta^2 + \left({\kappa_{B,0}}/{2} \right)^2 \right)\times\nonumber\\&&
\left(
\left( \Delta + \mu - \zeta_2 m^2 \right)^2 + \left({\kappa_{A,m}} \right)^2\right). \label{eqn:opo_P_th_l}
\end{eqnarray}
Here we've assumed again that the losses for the $\pm m$ modes are equal. Equation~(\ref{eqn:opo_P_th_l}) shows that we can use the modal detuning $\mu$ and the driving detuning $\Delta$ to select which modes reach threshold first and oscillate as the power is increased. Assuming that $g_{0,-mm}$ does not change significantly with the mode number, we see that for on-resonant driving $\Delta=0$, a minimum threshold can be achieved when $\mu=\zeta_2 m^2$, as long as $\mu$ and $\zeta_2$ have the same sign. In our case, the waveguide has normal dispersion, so $\zeta_2$ is negative, and we have roughly $\zeta_2/2\pi= -100~\textrm{kHz}$. The relation $m \approx \sqrt{\mu/\zeta_2}$ shows that the mode number selected is very sensitive to the modal detuning (set by temperature) which makes the degenerate oscillation mode challenging to obtain in a system with a large resonator and therefore very small $\zeta_2$ mode-spacing dispersion parameter. 

Interestingly, if the modal detuning $\mu$ is held constant while the pump detuning $\Delta$ is swept, the oscillation threshold can select very different modes $m$ with only small changes in $\Delta$.  When the laser is nearly resonant with $\Omega_0$, so $\Delta$ is small compared to the $B$ mode linewidth, the first term in parenthesis in eqn~(\ref{eqn:opo_P_th_l}) is minimized and does not vary strongly with detuning, while the second term is minimized whenever $\Delta+\mu \approx \zeta_2 m^2$.  This means that with a fixed laser input power, sweeping the laser across the second harmonic mode causes oscillation at very different mode numbers and explains the spectrum in figure~\ref{fig:3}b. For example, if we set $\Delta \ll \kappa_{B,0}$, we would obtain an approximate equation for the oscillating mode index (which should be rounded to obtain an integer, and requires $\mu+\Delta$ to have the same sign as $\zeta_2$):
\begin{eqnarray}
m \approx \sqrt{\frac{\mu+\Delta}{\zeta_2 }}
\label{eqn:muCondition}
\end{eqnarray}

For the real device, we observe disorder in the loss rates for different signal modes, which can result from fabrication imperfections or coupler dispersion. We can account for that in our threshold calculation
\begin{eqnarray}
P_{\text{th},m} &=& \frac{\hbar \Omega_0}{16 |g_{0,-mm}|^2}\frac{1 }{\kappa^{\text{(e)}}_{B,0}} \left( \Delta^2 + \left({\kappa_{B,0}}/{2} \right)^2 \right)\times\nonumber\\&&
\left(
\left( \Delta + \mu - \zeta_2 m^2 \right)^2 + \kappa_{A,m} \kappa_{A,-m}\right). \label{eqn:opo_P_th_l_disorder}
\end{eqnarray}
To obtain a relation for the OPO power output, we solve equations (\ref{eqn:dAdt}) - (\ref{eqn:dBdt}) for specific modes in steady-state. For the zero detuning of the pump mode $\Delta = 0$ and assuming $\mu = \zeta_2 m^2$, we have:
\begin{eqnarray}
A_m &=& 
\cfrac{4ig_{0,-mm} A_{-m}^{\ast} B_{0}}{\kappa_{A,m}}
\label{eqn:Am}\\
B_0 &=& 
\cfrac{2ig^{\ast}_{0,-mm}A_m A_{-m} + \sqrt{\kappa_{B,0}^{\text{(e)}}} G_{0} }{\kappa_{B,0}/2}
\label{eqn:Bk}.
\end{eqnarray}
Now, if we note that the oscillating amplitudes and coupling rate are complex $A_m = |A_m|\text{exp}(i\theta_m)$, $g_{k,00} = |g_{k,00}|\text{exp}(i\varphi)$, we can substitute equation (\ref{eqn:Am}) to (\ref{eqn:Bk}) and obtain
\begin{eqnarray}\cfrac{\kappa_{A,m} \kappa_{B,0} }{8i|g_{0,-mm}|^2} \cfrac{|A_{m}|}{|A_{-m}|} -2i|A_{m}||A_{-m}| + \nonumber\\
\cfrac{\sqrt{\kappa_{B,0}^{\text{(e)}}} }{|g_{0,-mm}|} G_{0} e^{i(\varphi - \theta_m - \theta_{-m})} = 0.
\label{eqn:ImCondition}
\end{eqnarray}
This requires the exponential $\text{exp}(i(\varphi - \theta_m - \theta_{-m}))$ to be purely imaginary, $\varphi - \theta_m - \theta_{-m} = \pi/2 + d\cdot\pi$, where $d\in\mathbb{Z}$. This phase relation shows that the sum of the phases of the OPO output are locked to the phase of the pump. As a result, we can use equation (\ref{eqn:Am}) to find that
\begin{eqnarray}
\cfrac{|A_m|}{|A_{-m}|} = \sqrt{\cfrac{\kappa_{A,-m}}{\kappa_{A,m}}},
\label{eqn:AmAnrelation}
\end{eqnarray}
and solve equation (\ref{eqn:ImCondition}) for the photon flux of both signal modes of the OPO:
\begin{eqnarray}
|A_{m}|^2 = \sqrt{\cfrac{\kappa_{A,-m}}{\kappa_{A,m}}} \cfrac{\sqrt{\kappa_{B,0}^{\text{(e)}}} G_{0}}{2|g_{0,-mm}|}
-\cfrac{\kappa_{B,0} \kappa_{A,m}}{16 |g_{0,-mm}|^2}
\label{eqn:AmModSq}
\\
|A_{-m}|^2 = \sqrt{\cfrac{\kappa_{A,m}}{\kappa_{A,-m}}} \cfrac{\sqrt{\kappa_{B,0}^{\text{(e)}}} G_{0}}{2|g_{0,-mm}|}
-\cfrac{\kappa_{B,0} \kappa_{A,-m}}{16 |g_{0,-mm}|^2}.
\label{eqn:AnModSq}
\end{eqnarray}
To analyze the total output power of the OPO in experiment, we sum over the power of two signal modes
\begin{eqnarray}
P_{\text{out}} = 
\cfrac{4\eta_{B,0}}{\Omega_{0}}
\left(
\eta_{A,m}\omega_{m}+ 
\eta_{A,-m}\omega_{-m}
\right)
\times \nonumber \\
P_{\text{th,m}}
\left(
\sqrt{\cfrac{P_{\text{B,0}}}{P_{\text{th,m}}}} - 1
\right),
\label{eqn:PoutOPOwithDisorder}
\end{eqnarray}
with $\eta_k=\kappa^{\text{(e)}}_{k,0}/\kappa_{k,0}$ for $k=A,B$ being the cavity-waveguide coupling efficiency. $P_{\text{B,0}}$ is the pump power of the SH mode and $P_{\text{th,m}}$ is a generalized OPO threshold, which includes disorder in the total loss rates of fundamental modes:
\begin{eqnarray}
P_{\text{th,m}} &=&
\cfrac{\hbar \Omega_{0}}{64|g_{0,-mm}|^2} \cfrac{\kappa_{B,0}^2\kappa_{A,m}\kappa_{A,-m}}{\kappa_{B,0}^{\text{(e)}}} = 
\nonumber\\&&
\cfrac{\hbar \Omega_{0}}{16 \eta_B} \cfrac{\sqrt{ \kappa_{A,m} \kappa_{A,-m}}}{C_{0,m} },
\label{eqn:PthresholdWithDisorder}
\end{eqnarray}
where $C_{0,m} \equiv 4|g_{0,-mm}|^2/\left(\sqrt{\kappa_{A,m}\kappa_{A,-m}}\kappa_{B,0}\right)$ is the vacuum cooperativity for $m^{\text{th}}$ pair of signal modes. Note that this relation agrees with equation (\ref{eqn:opo_P_th_l_disorder}) for the case of modal and laser detuning optimized for $m^{\text{th} }$ OPO sideband.

\subsection{Second Harmonic Generation Efficiency}

Starting from the coupled mode equations~(\ref{eqn:dAdt}) and (\ref{eqn:dBdt}), we now assume that only $A_0$ is excited, i.e., we are driving the mode at $\omega_0$ and all other mode FH amplitudes are $0$:
\begin{eqnarray}
\frac{\textrm{d}}{\textrm{d}t}  A_0 &=& - \frac{\kappa_{A,0}}{2} A_0 - 2 i  \sum_{k} g_{k,00} A_0^\ast B_k e^{-i\delta_{k,00}t} \nonumber \\&&~~~~~~~~~ - \sqrt{\kappa^{\text{(e)}}_{A,0}}F_0 e^{-i(\omega_L - \omega_0)t} \\
\frac{\textrm{d}}{\textrm{d}t}  B_k &=& -\frac{\kappa_{B,k}}{2} B_k - i g^{\ast}_{k,00} A_0^2 e^{+i\delta_{k, 00}t}.
\end{eqnarray}
To solve these equations, we go into a frame that rotates with the laser detuning frequency $\delta =\omega_L - \omega_0$, so $ A_0 = \Tilde A_0 e^{-i\delta t}$, $B_k = \Tilde B_k e^{-i(2\delta-\delta_{k, 0 0})t} = \Tilde B_k e^{-i(2\omega_L - \Omega_k)t}$:
\begin{eqnarray}
\frac{\textrm{d}}{\textrm{d}t}  \Tilde A_0 &=& \left(i\delta-\frac{\kappa_{A,0}}{2}\right) \Tilde A_0 - 2 i  \sum_{k} g_{k,00} \Tilde A_0^\ast \Tilde B_k - \nonumber\\&&~~~~~~~~~ \sqrt{\kappa^{\text{(e)}}_{A,0}}F_0 \\
\frac{\textrm{d}}{\textrm{d}t}  \Tilde B_k &=& \left(i(2\omega_L-\Omega_k)-\frac{\kappa_{B,k}}{2}\right) \Tilde B_{k} - i  g_{k,00}^\ast \Tilde A_0^2 .
\end{eqnarray}
We can solve these in steady state to obtain:
\begin{eqnarray}
\Tilde B_k = \frac{ig^{\ast}_{k,00} \Tilde A_0^2}{i(2\omega_L-\Omega_k)-\frac{\kappa_{B,k}}{2}}\label{eqn:steadystate_BK}
\end{eqnarray}
\begin{eqnarray}
0 &=& \left( i\delta - \frac{\kappa_{A,0}}{2}\right) \Tilde A_0 +  \sum_{k} \frac{2 |g_{k,00}|^2 |\Tilde A_0|^2}{i(2\omega_L - \Omega_k)-\frac{\kappa_{B,k}}{2}} \Tilde A_0 -\nonumber\\&&~~~~~~~~~ \sqrt{\kappa^{\text{(e)}}_{A,0}} F_0 \label{eqn:steadystate_A0}
\end{eqnarray}
There are a couple of interesting things to note about the last equation. Note that each SH mode at $\Omega_k$ contributes effective nonlinear loss and detuning terms to the FH mode at $\omega_0$:
\begin{eqnarray}
\text{Detuning:}~&& - \sum_k \frac{2(2\omega_L - \Omega_k)}{(2\omega_L - \Omega_k)^2 + \left(\frac{\kappa_{B,k}}{2}\right)^2} g_{k,00}^2 |A_0|^2 \label{eqn:2photon_detuning}\\
\text{Loss:}~&& \sum_k \frac{2\kappa_{B,k}}{(2\omega_L - \Omega_k)^2 + \left(\frac{\kappa_{B,k}}{2}\right)^2} g_{k,00}^2 |A_0|^2 \label{eqn:2photon_loss}
\end{eqnarray}
For large $2\omega_L - \Omega_{k} \gg \kappa_b$, we see an effect which is primarily a frequency shift and looks much like a $\chi^{(3)}$ cavity frequency shift.%

From here on, we assume that only one SH mode $(k=0)$ is significantly excited. The photons generated at the $B_0$ mode frequency are emitted from the device generating a photon flux $|G_{\text{out},0}|^2$ at the SH frequency where $G_{\text{out},0} = \sqrt{\kappa^{\text{(e)}}_{B,0}} \Tilde B_0$. To find $\Tilde B_0$, we need to calculate $\Tilde A_0$ (eqn~\ref{eqn:steadystate_BK}), which is given implicitly by 
\begin{eqnarray}
0 &=& \left( i\delta - \frac{\kappa_{A,0}}{2}\right) \Tilde A_0 +  \frac{ 2|g_{0,00}|^2 |\Tilde A_0|^2}{i(2\omega_L - \Omega_0)-\frac{\kappa_{B,0}}{2}} \Tilde A_0 -\nonumber\\&&~~~~~~~~~ \sqrt{\kappa^{\text{(e)}}_{A,0}} F_0. \label{eqn:steadystate_A0_single_mode}
\end{eqnarray}
For the fits shown in the paper, this equation was solved numerically. Here we assume $\delta=0$ and approximate the solutions in two limits, 1) the low-power limit where the first term is dominant, and 2) the high-power limit where the second term is dominant. The cross-over between these two limits occurs at
\begin{eqnarray}
2 C_0 n_A^{(0)} = 1
\end{eqnarray}
where $C_0 = 4|g_{0,00}|^2/\kappa_{A,0}\kappa_{B,0}$ is a cooperativity parameter and $n_A^{(0)} = {4 \kappa^{\text{(e)}}_{A,0}|F_0|^2 }/{{\kappa_{A,0}^2}} $ is the number of intracavity photons which would be excited in the absence of nonlinearity. 
Solving the above equation in the two limits gives us 
\begin{eqnarray}
|\Tilde A_0|^2 = n_A^{(0)}, \textrm{and}~~~~
|\Tilde A_0|^2 = \left(\frac{n_A^{(0)}}{4 C_0^2}\right)^{1/3}\nonumber
\end{eqnarray}
for the low- and high-power limits respectively. We define the second harmonic generation power efficiency 
\begin{eqnarray}
\eta_{\text{SHG}} \equiv \frac{P_\text{out}}{P_\text{in}} = \frac{2|G_{\text{out},0}|^2}{|F_0|^2},
\end{eqnarray}
which after some manipulation, can be written in terms of $|\Tilde A_0|^2$:
\begin{eqnarray}
\eta_{\text{SHG}} &=& \frac{8 \eta_A\eta_B C_0}{n_A^{(0)}} |\Tilde A_0|^4\nonumber\\
&=& \begin{cases}
      8 \eta_A \eta_B C_0 n^{(0)}_A & \text{low power}\\
      \cfrac{4 \eta_A \eta_B}{(2n^{(0)}_A C_0)^{1/3}}& \text{high power}
    \end{cases}       
\end{eqnarray}
It is apparent that at low power, the efficiency increases linearly, but is then saturated at high power. This can be understood from an impedance matching perspective. As the pump power is increased, the FH cavity resonance senses a two-photon loss proportional to $8|g_{0,00} \Tilde A_0|^2/\kappa_{B,0}$ (see eqn.~\ref{eqn:2photon_loss}). As this loss starts to exceed the cavity linewidth, its effective coupling rate to the waveguide is reduced, preventing input light from coupling efficiently into the cavity to be frequency-doubled. At very high power, the efficiency actually begins to go down as  $P^{-1/3}$. The model assumes that only the $\Tilde B_0$ and $\Tilde A_0$ modes are excited. As we saw in the case of a directly driven OPO, at sufficiently large $\Tilde B_0$, $\Tilde A_{\pm m}$ start to oscillate, which causes this model to break down and the system to go into cascaded optical parametric oscillation.

\subsection*{Cascaded Optical Parametric Oscillation}
 
Consider the same driving as in the previous section, where a laser drive at the fundamental with frequency $\omega_L$ excites $\Tilde A_0$ and generates an intracavity population in the second harmonic mode $\Tilde B_0$. From the section on the oscillation threshold, we know that at a sufficiently value of $|\Tilde B_0|$, the equations of motion for mode amplitudes $\Tilde A_{\pm m}$ become unstable and set of oscillations, with a threshold condition given by an equation very similar to eqn.~(\ref{eqn:opo_stab_crit_1}):
\begin{eqnarray}
16|g_{0,-mm}|^2 |\Tilde B_0|^2  \geq \left(2\delta +\mu - {\zeta_2}m^2\right)^2 +  {\kappa_{A,m}}{\kappa_{A,-m}}.\nonumber \label{eqn:copo_stab_crit_3}
\end{eqnarray}
We call this a cascaded OPO, since a cascade of two back-to-back $\chi^{(2)}$ processes leads to a an effective $\chi^{(3)}$ four-wave mixing process. It is clear from the oscillation condition that the threshold is highly detuning-dependent, and also depends on the dispersion parameter $\zeta_2$. In our case, $\zeta_2/2\pi\approx -100~\textrm{kHz}$ and so the oscillation threshold is lower with the laser tuned to the red side ($\delta<0$) when the modal detuning $\mu \approx 0$.

\subsection{Nonlinear Coupling Rate}\label{sec:g_theory}

We derive the nonlinear coupling rate from the interaction energy density in the three-wave mixing process. Given the electric field distribution $\m E{} = (E^x,E^y,E^z)$ The  interaction energy density  is given by:
\begin{eqnarray}
U_{\chi^{(2)}} =
\cfrac{\varepsilon_{0}}{3} \sum_{\alpha \beta \gamma} \chi_{\alpha \beta \gamma}^{(2)} {E}^{\alpha} {E}^{\beta} {E}^{\gamma},
\label{eqn:nlInteractionE}
\end{eqnarray}
each of the three waves can be expressed using spatial complex amplitudes $\m Em, \m En, \m Ek$ as follows:
\begin{eqnarray}
\m{E}{} &=& 
A_m \m {E}m e^{-i\omega_m t} +
A_n \m{E}n e^{-i\omega_n t} + B_k\m{E}k e^{-i\Omega_k t} \nonumber\\&&~~~~~~~~~~ + \textrm{h.c.}
\label{eqn:Ealpha}
\end{eqnarray}
To calculate the nonlinear coupling rate we focus on three specific modes in the sum and evaluate equation \ref{eqn:nlInteractionE} by averaging away the rapidly rotating terms. It selects only energy-conserving terms of the sum. Since the second-order nonlinear tensor has a full permutation symmetry, for the non-degenerate we find that
\begin{eqnarray}
U^{k,nm}_{\chi^{(2)}} &=&
2\varepsilon_{0} \sum_{\alpha \beta \gamma}
\chi^{(2)}_{\alpha \beta \gamma}
\left(
{E}^{\alpha\ast}_{k} {E}^{\beta}_{m} {E}^{\gamma}_{n} B_k^{\ast} A_m A_n  + \text{h.c.} \right)
= \nonumber \\&&
2\varepsilon_{0}
\left(
\m{E}k^{\ast} \bar{\bar{\chi}}^{(2)}:\m{E}{m} \m{E}{n} B_k^{\ast} A_m A_n  + \text{h.c.}
\right).
\label{eqn:nlInteractionE_2}
\end{eqnarray}
Integrating over this energy density gives us the total energy of the system, which we use to derive the equations of motion (\ref{eqn:dAdt}) and (\ref{eqn:dBdt}). We choose normalization of the modal field $\m {E}{k}$ so that the total energy corresponding to an amplitude $A_k$ is $\hbar \omega_k |A_k|^2$. More precisely, given unitless field profiles $\m ei$ (with $\textrm{max} (\m ei) = 1$), we introduce normalization factors $N_i$, defined by $\m Ei = N_i \m ei$. The energy condition then fixes these normalization factors as
\begin{eqnarray}
N_i &=& \sqrt{\cfrac{\hbar \omega_i}{2\int \m ei^\ast \bar{\bar{\varepsilon}}(\m r{})\m ei dV}} \nonumber\\
 &=& \sqrt{\cfrac{\hbar \omega_i}{2\varepsilon_0 L \int  \m ei^\ast \bar{\bar{\varepsilon}}_r(\m r{})\m ei dA}}\nonumber\\
 &=& \sqrt{\cfrac{\hbar \omega_i}{2\varepsilon_0 L \bar n_i^2}}\frac{1}{\sqrt{\mathcal{A}_i}}
\label{eqn:nlInteractionE_norm}
\end{eqnarray}
Here, we introduced the effective mode area for each mode as $\mathcal{A}_i = \int_{A} |\m{e}{i}|^2 dA$, and define the average index as $ \bar n_i^2=\int  \m ei^\ast \bar{\bar{\varepsilon}}_r(\m r{})\m ei dA /\mathcal{A}_i$. To find the energy, we integrate equation \ref{eqn:nlInteractionE_2} over the mode volume. We account for a partially-poled racetrack resonator by introducing the poled length fraction $\lambda$ as a ratio of the poled region to the total resonator length $L$. The final expression for the nonlinear coupling rate is given by:
\begin{eqnarray}
g_{k,nm} = \cfrac{\lambda}{\sqrt{2} \pi} \sqrt{\cfrac{\hbar\omega_m\omega_n\Omega_k}{\varepsilon_0 L \bar n_k^2 \bar n_m^2 \bar n_n^2}} \cfrac{\mathcal{O}}{\sqrt{\mathcal{A}_m \mathcal{A}_n \mathcal{A}_k}},
\label{eqn:nlCouplingRate}
\end{eqnarray}
where $\mathcal{O}$ represents the mode overlap integral over the waveguide cross section area
\begin{eqnarray}
\mathcal{O} = 
\int_{A} \m{e}{k}^{\ast} \bar{\bar{\chi}}^{(2)}:
\m{e}{m} \m{e}{n} dA.
\label{eqn:overlapIntegral}
\end{eqnarray}
For our numerical waveguide calculations we use a finite-element mode solver (COMSOL).

\subsection{Numerical Simulations of Dynamics}

We numerically integrate the coupled mode differential equations (\ref{eqn:dAdt}) and (\ref{eqn:dBdt}) to understand how the transmission spectra change when the system starts to oscillate and how disorder affects the emission spectra. We integrate the coupled mode equations with with $181$ $A$ modes and $31$ $B$ modes for $600$ nanoseconds which is sufficiently long for the system to stabilize. We use the measured parameters from the SHG experiment for the $\omega_0$ and $\Omega_0$ modes, and assume that the other modes are spaced by the measured FSR (which agrees with the theory prediction) and have the same quality factors. The resulting spectra for are shown in Extended Figure Fig. \ref{fig:FigSI2}. We then perform the same simulation but with the quality factors and detunings of the other modes now having disorder (normally distributed fluctuations of total $Q$ and mode frequency) on the order of $1\%$ (Fig. \ref{fig:FigSI3}) and $10\%$ (Fig. \ref{fig:FigSI4}) of the cavity linewidth.

\section*{Acknowledgments}
The authors wish to thank NTT Research for their financial and technical support. This work was funded by the U.S. Department of Defense through the DARPA Young Faculty Award (YFA), the DARPA LUMOS program (both supported by Dr. Gordon Keeler), and through the U.S. Department of Energy through Grant No. DE-AC02-76SF00515 (through SLAC). Part of this work was performed at the Stanford Nano Shared Facilities (SNSF), supported by the U.S. National Science Foundation under award ECCS-2026822. H.S.S. acknowledges support from the Urbanek Family Fellowship. J.F.H. was supported by the National Science Foundation Graduate Research Fellowship Program. V.A. was supported by the Stanford Q-FARM Bloch Fellowship Program. A.-H.S.N. acknowledges the David and Lucille Packard Fellowship, and the Stanford University Terman Fellowship.

\section*{Author contributions}
T.P.M. and H.S.S. designed the device.
T.P.M, H.S.S., and V.A. fabricated the device.
T.P.M., H.S.S., V.A., J.M., C.J.S., J.F.H. and C.L. developed the fabrication process.
V.A., M.J., M.M.F. and A.H.S.-N. provided experimental and theoretical support.
T.P.M. and H.S.S. performed the experiments and analysed the data. A.H.S.-N performed numerical simulations.
T.P.M., H.S.S. and A.H.S.-N. wrote the manuscript.  
T.P.M, H.S.S. and A.H.S.-N. conceived the experiment, and A.H.S.-N. supervised all efforts. 

\section*{Author information}

The authors declare no competing financial interests. All correspondence should be addressed to A. H. Safavi-Naeini (safavi@stanford.edu).

\begin{figure*}[h] 
  \centering
  \includegraphics[width=\textwidth]{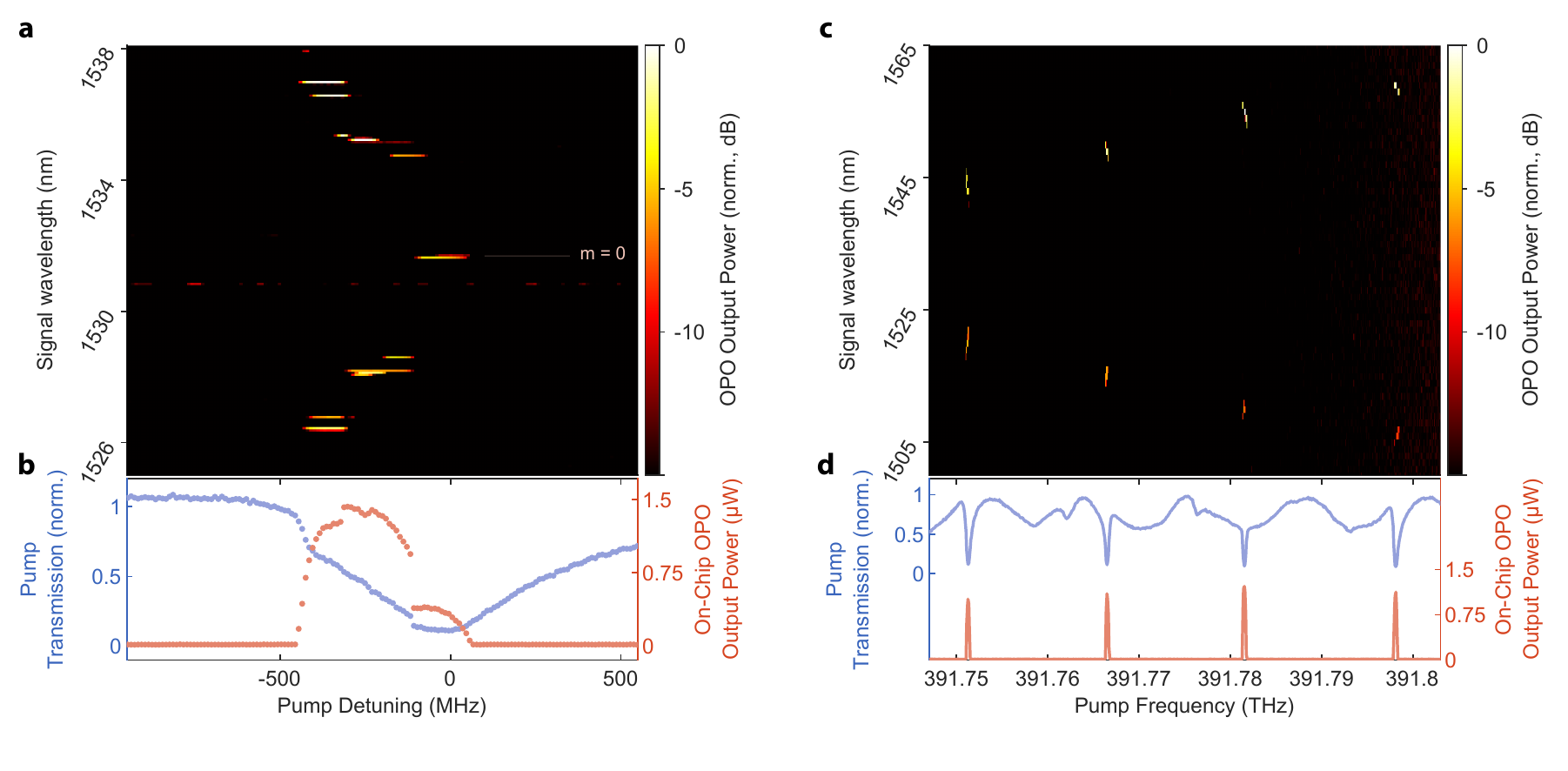}
\caption{\textbf{Extended Data Figure: Optical Parametric Oscillation.} 
\textbf{a}, Tuning of the OPO close to $\mu=0$, at a pump power of 250~$\mu\text{W}$. We observe degenerate parametric oscillation at 1531.7~nm ($m=0$) and nondegenerate operation for blue-detuning of the pump laser. \textbf{b}, SH resonance lineshape (blue points) aligned with the OPO response (red points) collected above threshold with 250~$\mu\text{W}$ of on-chip pump power. The distinct feature at zero-detuning corresponds to degenerate oscillation.
\textbf{c}, Sweeping the pump laser over four neighboring modes shows that the signal/idler pair center frequencies are different for each OPO. This is due to the different modal detuning $\mu=0$ experienced by each OPO due to the difference in dispersion at 765 nm and 1530 nm. The higher frequency OPOs have larger $|\mu|$ (see discussion in Section G of Methods).  
This shows that additional tuning range of the device's output frequency can be extended to about 2.75 THz by utilizing multiple OPOs of a single resonator while keeping the chip temperature fixed.  
\textbf{d}, SH resonance lineshape (blue line) aligned with the OPO response (red line) collected above threshold at 250~$\mu\text{W}$ of pump power on chip.
}
\label{fig:FigSI1}
\end{figure*}

\begin{figure*}[h] 
  \centering
  \includegraphics[width=\textwidth]{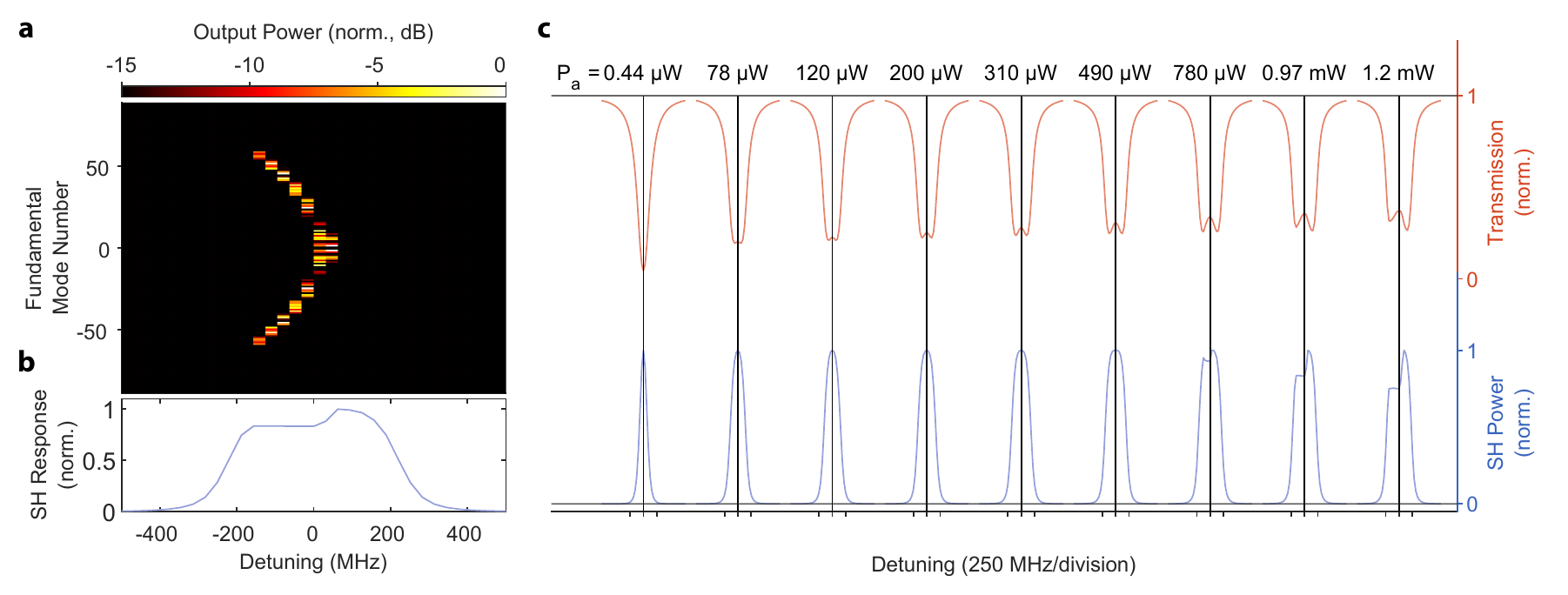}
\caption{\textbf{Simulated Effect of 0.1\% Disorder in the  Mode-to-Mode Loss Rates on the Cascaded OPO.} 
\textbf{a}, Cascaded OPO signal as a function of pump detuning, simulated at 0.97 mW of pump power.
\textbf{b}, Second Harmonic lineshape corresponding to the cascaded OPO in panel \textbf{a}.
\textbf{c}, Lineshape evolution as a function of power, asymmetry is developed below 780~$\mu\text{W}$, which agrees with the experiment.
}
\label{fig:FigSI2}
\end{figure*}

\begin{figure*}[h] 
  \centering
  \includegraphics[width=\textwidth]{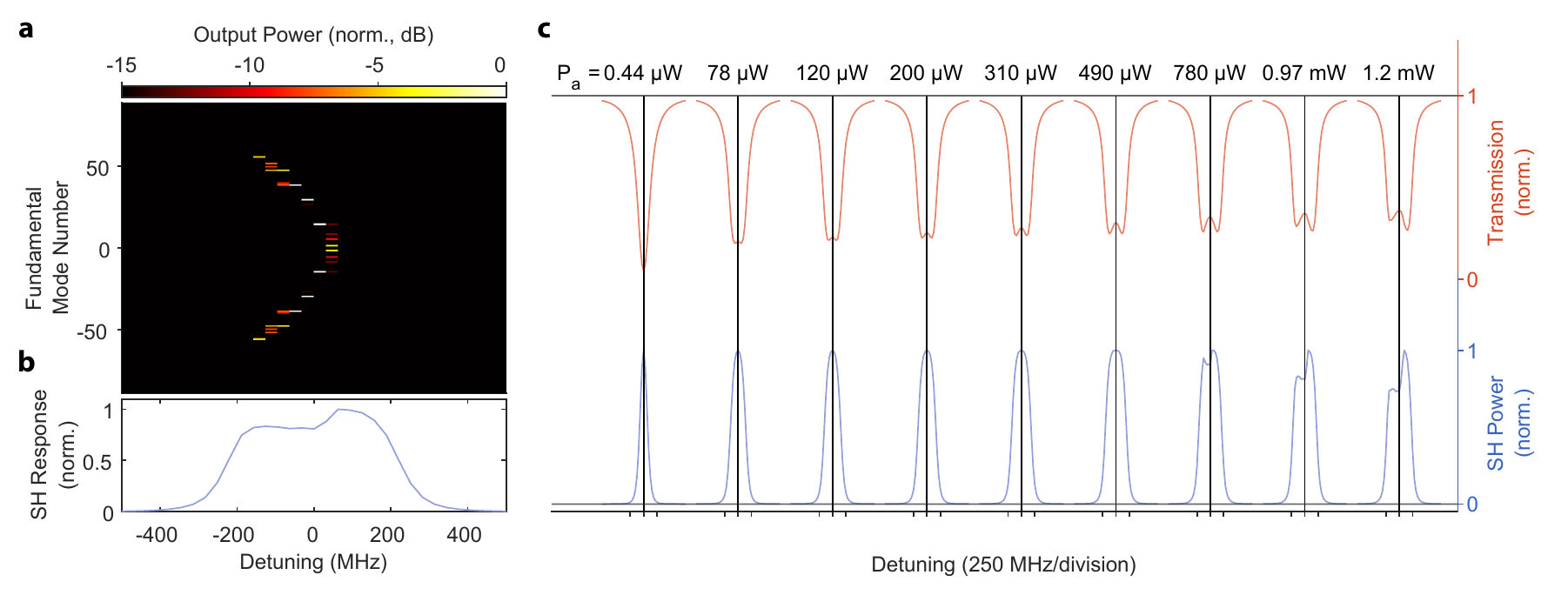}
\caption{\textbf{Simulated Effect of 1\% Disorder in the  Mode-to-Mode Loss Rates on the Cascaded OPO.} 
\textbf{a}, Cascaded OPO signal as a function of pump detuning, simulated at 0.97 mW of pump power.
\textbf{b}, Second Harmonic lineshape corresponding to the cascaded OPO in panel \textbf{a}.
\textbf{c}, Lineshape evolution as a function of power, asymmetry is developed below 780~$\mu\text{W}$, which agrees with the experiment.
}
\label{fig:FigSI3}
\end{figure*}

\begin{figure*}[h] 
  \centering
  \includegraphics[width=\textwidth]{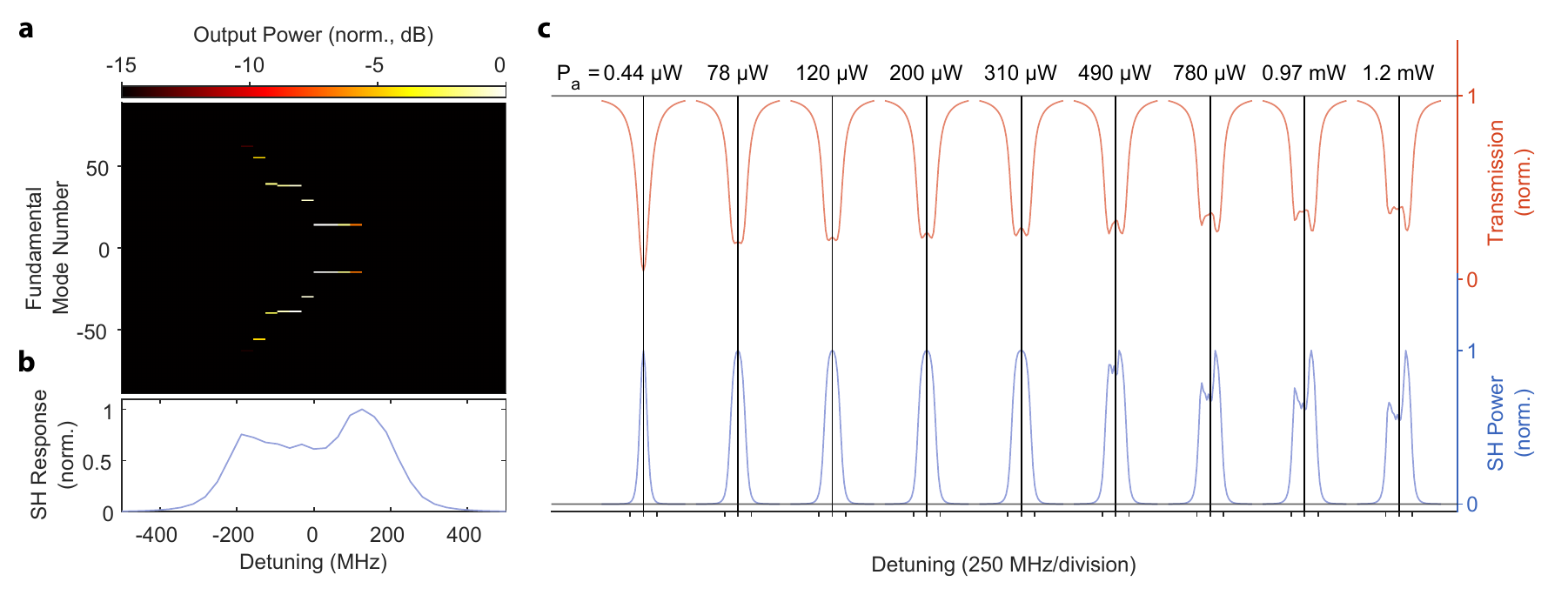}
\caption{\textbf{Simulated Effect of 10\% Disorder in the  Mode-to-Mode Loss Rates on the Cascaded OPO.} 
\textbf{a}, Cascaded OPO signal as a function of pump detuning, simulated at 0.97 mW of pump power.
\textbf{b}, Second Harmonic lineshape corresponding to the cascaded OPO in panel \textbf{a}.
\textbf{c}, Lineshape evolution as a function of power, asymmetry is developed around 490~$\mu\text{W}$, lower than in the experiment.
}
\label{fig:FigSI4}
\end{figure*}

\begingroup
\setlength{\tabcolsep}{3.5pt} %
\renewcommand{\arraystretch}{1.1} %
\begin{table*}[h]
\begin{tabular}{c|c|c|c|c|c|c|c|c|c}
Device & $\lambda_{A}\,(\text{nm})$ & $\lambda_{B}\,(\text{nm})$ & $Q_{A}\,(10^{6})$ & $Q_{A}^{\text{(i)}}\,(10^{6})$ & $Q_{B}\,(10^{6})$ & $Q_{B}^{\text{(i)}}\,(10^{6})$ & $g_{0,nm}\,(\text{kHz})$ & $\eta_\text{FH}\, (\%)$ & $\eta_\text{SH}\, (\%)$ \\ \hline\hline
 
OPO & \begin{tabular}{@{}c@{}}$\lambda_{A,m} = 1521.05 $ \\  $\lambda_{A,-m} = 1542.43 $\end{tabular}  
& 765.77 & 
\begin{tabular}{@{}c@{}} $Q_{A,m}=0.68$  \\  $Q_{A,-m}=0.94$  \end{tabular}
& 
\begin{tabular}{@{}c@{}} $Q_{A,m}^{\text{(i)}}=0.80$  \\  $Q_{A,-m}^{\text{(i)}}=1.50$  \end{tabular} 
& 0.88 & 1.50 & 150 & 37 & 13 \\ \rule{0pt}{4ex}  
SHG & 1549.40 & 774.70 & 0.74 & 1.2 & 0.82 & 1.2 & 130 & 26 & 11
\end{tabular}
\caption{
\textbf{Summary of the Measured Device Parameters.}
We summarize the wavelengths ($\lambda_{A}$ and $\lambda_{B}$), total ($Q$) and internal ($Q^{\text{(i)}}$) quality factors of all of the resonances used in our OPO and SHG experiments. We list the nonlinear coupling factors ($g_{0,nm}$) and edge coupling efficiencies at both wavelengths ($\eta_\text{FH}$ and $\eta_\text{SH}$) for particular devices.
}
\end{table*}
\endgroup

\end{document}